\documentclass[conference,final,10pt,a4paper]{IEEEtran}

\usepackage{cite}
\usepackage[pdftex]{graphicx}
 \graphicspath{{../pdf/}{../jpeg/}}
 \DeclareGraphicsExtensions{.pdf,.jpeg,.png}
\usepackage{pgf}
\usepackage{pgfplots}
\usepackage[cmex10]{amsmath}
\usepackage{amssymb}
\usepackage{algorithmic}
\usepackage{array}
\usepackage{mdwmath}
\usepackage{mdwtab}
\usepackage{url}
\usepackage{fixedmultirow}
\usepackage{trfsigns}
\usepackage{caption}
\usepackage{subcaption}
\usepackage{fancyhdr}

\graphicspath{{graphics/}}
\graphicspath{{../Bilder/}}

\usetikzlibrary{calc,trees,positioning,arrows,chains,shapes.geometric,%
 decorations.pathreplacing,decorations.pathmorphing,shapes,%
 matrix,shapes.symbols,plotmarks,decorations.markings,shadows,fit,patterns}
\usetikzlibrary{spy,backgrounds}
\usepgfplotslibrary{groupplots}

\pgfplotsset{compat=1.3}
\tikzset{every picture/.style={>=latex}} 
\tikzset{every tick label/.style={font=\footnotesize}} 
\tikzset{every axis label/.style={font=\small}} 
\pgfsetplotmarksize{2pt}

\tikzset{orientation/.is choice,
    orientation/lr/.style={anchor=west,right=1},
    orientation/lr2/.style={anchor=west,right=2},
    orientation/lrd/.style={anchor=west,below=1},
    orientation/lrd2/.style={anchor=west,below=2},
    orientation/rl/.style={anchor=east,left=1},
    orientation/rl2/.style={anchor=east,left=2},
    orientation/ud/.style={anchor=north,below=1},
    orientation/du/.style={anchor=south,above=1},
    orientation/rld/.style={anchor=east,below=1},
    orientation/rld2/.style={anchor=east,below=2},
}
\tikzstyle{scare} = [
]
\tikzstyle{syslinear} = [
 drop shadow={shadow xshift=.6mm,
              shadow yshift=-.6mm},
 fill=white,
 anchor=west,
 rectangle,
 draw=black,
 minimum height=5mm,
 minimum width=5mm,
 inner xsep=0.5em
]

\tikzstyle{sysnonlinear} = [
 drop shadow={shadow xshift=.8mm,
              shadow yshift=-.8mm},
 fill=white,
 double,
 anchor=west,
 rectangle,
 draw=black,
 minimum height=5mm,
 minimum width=5mm,
 inner xsep=0.5em
]

\tikzstyle{syssource} = [
 anchor=west,
 ellipse,
 draw=black,
 minimum height=1.5ex,
 minimum width=1.5em,
 inner xsep=0.5em
]

\tikzstyle{syssink} = [
 anchor=west,
 ellipse,
 draw=black,
 minimum height=1.5ex,
 minimum width=1.5em,
 inner xsep=0.5em
]

\tikzstyle{syssplit} = [
 fill=black,
 draw=black,
]

\tikzstyle{sysadd} = [
 draw,circle,inner sep=-1pt,
]

\tikzstyle{sysmul} = [
 draw,circle,inner sep=-1pt,
]

\definecolor{MyHSBGreen}{hsb}{0.34065,1,0.91}
\pgfplotscreateplotcyclelist{colors4}{%
 {black!100!white},
 {black!75!white},
 {black!50!white},
 {black!25!white},
}
\pgfplotscreateplotcyclelist{colors5}{%
 {black!100!white},
 {black!75!white},
 {black!50!white},
 {black!25!white},
 {densely dashed, black!100!white},
 {densely dashed, black!75!white},
}
\pgfplotscreateplotcyclelist{colors10}{%
 {black!100!white},
 {black!75!white},
 {black!50!white},
 {black!25!white},
 {densely dashed, black!100!white},
 {densely dashed, black!75!white},
 {densely dashed, black!50!white},
 {densely dashed, black!25!white},
 {densely dotted, black!100!white},
 {densely dotted, black!75!white},
 {densely dotted, black!50!white},
}

\pgfplotscreateplotcyclelist{colorsBERMDDFSE}{%
                 every mark/.append style={fill=black},mark=o\\%
                 every mark/.append style={fill=black},mark=star\\%
                 every mark/.append style={fill=black},mark=triangle\\%
                 every mark/.append style={scale=.7,fill=black},mark=square\\%
                 every mark/.append style={fill=black},mark=diamond\\%
  densely dashed,every mark/.append style={solid},mark=pentagon*\\%
  densely dashed,every mark/.append style={solid},mark=*\\%
  densely dashed,every mark/.append style={solid},mark=triangle*\\%
  densely dashed,every mark/.append style={scale=.7,solid},mark=square*\\%
  densely dashed,every mark/.append style={solid},mark=diamond*\\%
  densely dashed,every mark/.append style={solid},mark=star*\\%
}
\pgfplotscreateplotcyclelist{colors1Empty4Full}{%
                 every mark/.append style={fill=black},mark=o\\%
  densely dashed,every mark/.append style={solid},mark=*\\%
  densely dashed,every mark/.append style={solid},mark=triangle*\\%
  densely dashed,every mark/.append style={scale=.7,solid},mark=square*\\%
  densely dashed,every mark/.append style={solid},mark=diamond*\\%
}

\pgfplotscreateplotcyclelist{colors4Empty4Full}{%
                 every mark/.append style={fill=black},mark=o\\%
                 every mark/.append style={fill=black},mark=triangle\\%
                 every mark/.append style={scale=.7,fill=black},mark=square\\%
                 every mark/.append style={fill=black},mark=diamond\\%
  densely dashed,every mark/.append style={solid},mark=*\\%
  densely dashed,every mark/.append style={solid},mark=triangle*\\%
  densely dashed,every mark/.append style={scale=.7,solid},mark=square*\\%
  densely dashed,every mark/.append style={solid},mark=diamond*\\%
}
\pgfplotscreateplotcyclelist{colors6Empty4Full}{%
                 every mark/.append style={fill=black},mark=o\\%
                 every mark/.append style={fill=black},mark=star\\%
                 every mark/.append style={fill=black},mark=triangle\\%
                 every mark/.append style={scale=.7,fill=black},mark=square\\%
                 every mark/.append style={fill=black},mark=diamond\\%
                 every mark/.append style={fill=black},mark=pentagon\\%
  densely dashed,every mark/.append style={solid},mark=*\\%
  densely dashed,every mark/.append style={solid},mark=triangle*\\%
  densely dashed,every mark/.append style={scale=.7,solid},mark=square*\\%
  densely dashed,every mark/.append style={solid},mark=diamond*\\%
}
\pgfplotscreateplotcyclelist{colors8Empty4Full}{%
                 every mark/.append style={fill=black},mark=o\\%
                 every mark/.append style={fill=black},mark=star\\%
                 every mark/.append style={fill=black},mark=triangle\\%
                 every mark/.append style={scale=.7,fill=black},mark=square\\%
                 every mark/.append style={fill=black},mark=diamond\\%
                 every mark/.append style={fill=black},mark=pentagon\\%
                 every mark/.append style={fill=black},mark=oplus\\%
                 every mark/.append style={fill=black},mark=otimes\\%
  densely dashed,every mark/.append style={solid},mark=*\\%
  densely dashed,every mark/.append style={solid},mark=triangle*\\%
  densely dashed,every mark/.append style={scale=.7,solid},mark=square*\\%
  densely dashed,every mark/.append style={solid},mark=diamond*\\%
}

\pgfplotscreateplotcyclelist{colors}{%
                 every mark/.append style={fill=black},mark=o\\%
                 every mark/.append style={fill=black},mark=star\\%
                 every mark/.append style={fill=black},mark=triangle\\%
                 every mark/.append style={scale=.7,fill=black},mark=square\\%
                 every mark/.append style={fill=black},mark=diamond\\%
                 every mark/.append style={fill=black},mark=pentagon\\%
                 every mark/.append style={fill=black},mark=oplus\\%
                 every mark/.append style={fill=black},mark=otimes\\%
                 every mark/.append style={fill=black},mark=asterisk\\%
                 every mark/.append style={fill=black},mark=+\\%
                 every mark/.append style={fill=black},mark=-\\%
                 every mark/.append style={fill=black},mark=|\\%
                 every mark/.append style={fill=black},mark=x\\%
                 every mark/.append style={fill=black},mark=*\\%
                 every mark/.append style={fill=black},mark=triangle*\\%
                 every mark/.append style={scale=.7,fill=black},mark=square*\\%
                 every mark/.append style={fill=black},mark=diamond*\\%
                 every mark/.append style={fill=black},mark=star*\\%
  densely dashed,every mark/.append style={solid,fill=gray},mark=o\\%
  densely dashed,every mark/.append style={solid,fill=gray},mark=triangle\\%
  densely dashed,every mark/.append style={scale=.7,solid,fill=gray},mark=square\\%
  densely dashed,every mark/.append style={solid,fill=gray},mark=diamond\\%
  densely dashed,every mark/.append style={solid,fill=gray},mark=star\\%
  densely dashed,every mark/.append style={solid,fill=gray},mark=pentagon*\\%
  densely dashed,every mark/.append style={solid,fill=gray},mark=*\\%
  densely dashed,every mark/.append style={solid,fill=gray},mark=triangle*\\%
  densely dashed,every mark/.append style={scale=.7,solid,fill=gray},mark=square*\\%
  densely dashed,every mark/.append style={solid,fill=gray},mark=diamond*\\%
  densely dashed,every mark/.append style={solid,fill=gray},mark=star*\\%
  mark=text,text mark=a\\%
  mark=text,text mark=b\\%
  mark=text,text mark=c\\%
  mark=text,text mark=d\\%
  mark=text,text mark=e\\%
  mark=text,text mark=f\\%
  mark=text,text mark=g\\%
}

\pgfdeclarelayer{background layer}
\pgfdeclarelayer{foreground layer}
\pgfsetlayers{background layer,main,foreground layer}

\newcommand{\ie}{\emph{i.e.}}
\newcommand{\eg}{\emph{e.g.}}
\newcommand{\cf}{\emph{cf.}}

\newcommand{\Eg}{\emph{E.g.}}

\providecommand{\de}[1]{\ensuremath{\mathop{\mathrm{d}}}}

\newcommand{\imag}{\ensuremath{\mathrm{j}}}

\newcommand{\entspricht}{\ensuremath{\mathrel{\widehat{=}}}} 

\IEEEoverridecommandlockouts
\title{Reduced Complexity Super-Trellis Decoding for Convolutionally Encoded Transmission Over ISI-Channels}
\author{
 \IEEEauthorblockN{Fabian~Schuh,
                   Andreas~Schenk, and
                   Johannes~B.~Huber}%
 \IEEEauthorblockA{Institute for Information Transmission,
                   Friedrich-Alexander-Universit\"at Erlangen-N\"urnberg, Germany\\ 
                   mail: \texttt{\{schuh,\,schenk,\,huber\}@LNT.de}}%
 \thanks{This work was supported by Federal Ministry of Economics and Technology
         (BMWi) within the project C-PMSE.}
}

\begin{document}
\maketitle
\begin{abstract}
 In this paper we propose a matched encoding (ME) scheme for convolutionally
 encoded transmission over intersymbol interference (usually called ISI)
 channels. A novel trellis description enables to perform equalization and
 decoding jointly, \ie, enables efficient super-trellis decoding. By means of
 this matched non-linear trellis description we can significantly reduce the
 number of states needed for the receiver-side Viterbi algorithm to
 perform maximum-likelihood sequence estimation. Further complexity reduction
 is achieved using the concept of reduced-state sequence estimation.
\end{abstract}
\begin{IEEEkeywords}
 ISI-channel;
 convolutionally encoded transmission;
 super-trellis decoding;
 matched decoding;
 reduced state sequence estimation;
 trellis-coded modulation;
\end{IEEEkeywords}
\IEEEpeerreviewmaketitle
\section{Introduction}

Convolutional coded pulse-amplitude modulation (PAM) is an attractive digital
communication scheme for transmission over intersymbol interference (ISI)
channels, when low latency is desired. Low latency, required \eg, for real-time
bidirectional communication, is obtained by the use of convolutional codes
(instead of block codes, \cf~\cite{LIT_tr_com_2009_hehn}) and dispense with
interleaving (as opposed to conventional bit-interleaved coded
modulation~\cite{141453}).

For this setup, the optimum receiver performs equalization of the ISI-channel
and decoding of the convolutional code jointly in a single
super-trellis~\cite{huber1992trelliscodierung}. This technique, however, is
commonly regarded prohibitively complex due to the large overall number of
states of a super-trellis. Hence, equalization and decoding are usually
performed subsequently in two separate processing steps, each based on its own
trellis description. As long as no interleaving can be applied, iterative
(Turbo-) decoding/equalization does not work satisfactorily.

In this paper, we merge the convolutional encoder and the ISI-channel into a
single non-linear trellis encoder with binary delay elements only. It is shown,
that the total number of states of this equivalent non-linear trellis
description is significantly smaller than the number of states in the usual
super-trellis. Consequently, this non-linear trellis description enables very
efficient implementation of optimum super-trellis decoding (STD) based on
maximum-likelihood sequence estimation (MLSE) using the Viterbi algorithm
(VA)~\cite{1054010} or other trellis-based decoding algorithms.

Combining this approach with reduced-state sequence estimation
(RSSE)~\cite{Spinnler1995,Eyuboglu1988,Eyuboglu1989} enables to further reduce
the computational complexity and thus offers a flexible trade-off between
performance (in terms of required signal-to-noise power ratio to guarantee a
target bit error rate) and receiver complexity.

The only requirement necessary for our approach is that the rate-$\frac{K}{n}$
convolutional code is matched to the $M$-ary modulation via $M=2^n$, \ie,
trellis-coded modulation (TCM). For sake of simplicity, we here consider
real-valued ASK only.

Note that recently, we have adopted a similar approach to receiver design of continuous
phase modulation (CPM) in combination with non-coherent differential
detection~\cite{Schu1301:Nonlinear}.

This paper is organized as follows: After the definition of the system model in
Sec.~\ref{sec:sysmodel}, we derive the equivalent non-linear trellis
description in Sec.~\ref{sec:matchedEncDerivation}. In
Sec.~\ref{sec:complexity} the reduced computational complexity is discussed and
Sec.~\ref{sec:RSSE} employs reduced-state sequence estimation (RSSE) for our
approach. The effectiveness of the proposed approach is validated by means of
numerical simulations in Sec.~\ref{sec:simresults}. The paper concludes with a
summary.

\section{System Model}\label{sec:sysmodel}

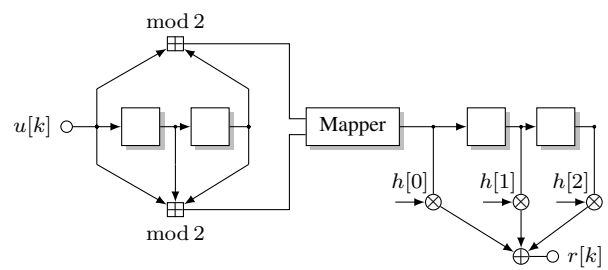
\begin{figure}[ht]\vspace{-1ex}
 \begin{center}
  \begin{tikzpicture}[>=latex,x=1em,y=4ex,font=\footnotesize,inner sep=0.3em,
                      node distance=10mm and 4mm]
   \node at (0,0) (u) {$u[k]$};
   \node[coordinate,right=of u] (in) {};
   \draw node[syslinear,right=of in] (T1) {};
   \draw node[syslinear,right=of T1] (T2) {};
   \draw node[sysadd, above=of T1.east,xshift=2mm,rectangle,inner xsep=-.3pt,inner ysep=-.2pt] (g1) {$+$};
   \draw node[sysadd, below=of T1.east,xshift=2mm,rectangle,inner xsep=-.3pt,inner ysep=-.2pt] (g2) {$+$};
   \node[anchor=south,at=(g1.north)] {$\operatorname{mod}2$};
   \node[anchor=north,at=(g2.south)] {$\operatorname{mod}2$};
   \path (u)       edge[o->] node[coordinate,pos=0.6] (u1) {} (T1)
         (T1)      edge[->]  node[coordinate,midway] (u2) {} (T2)
         (T2.east) edge[-]   node[coordinate,pos=1] (u3) {}  ++(2.5mm,0);
   \draw[fill]  (u1) circle(.5pt) -- ++(0,5mm) edge[->] (g1)
                (u3) circle(.5pt) -- ++(0,5mm) edge[->] (g1)
                (u1) circle(.5pt) -- ++(0,-5mm) edge[->] (g2)
                (u2) circle(.5pt) -- ++(0,-5mm) edge[->] (g2)
                (u3) circle(.5pt) -- ++(0,-5mm) edge[->] (g2);
   \node[syslinear,right=10mm,at=(T2.east)] (Mapper) {Mapper};
   \draw (g1) -- ++(15mm,0) |- ($(Mapper.west)+(0,3pt)$);
   \draw (g2) -- ++(15mm,0) |- ($(Mapper.west)-(0,3pt)$);
   \draw node[syslinear,at=(Mapper),xshift=15mm] (T3) {};
   \draw node[syslinear,right=of T3] (T4) {};
   \draw (Mapper)   edge[->] node[coordinate,midway] (u4) {} (T3)
         (T3)       edge[->] node[coordinate,midway] (u5) {} (T4)
         (T4.east)  edge[-] node[coordinate,pos=1]   (u6) {} ++(2.5mm,0);
   \node[sysmul,at=(u4),yshift=-10mm] (h1) {$\times$};
   \node[sysmul,at=(u5),yshift=-10mm] (h2) {$\times$};
   \node[sysmul,at=(u6),yshift=-10mm] (h3) {$\times$};
   \draw[<-] (h1) -- ++(-5mm,0) node[pos=0.5,anchor=south] {$h[0]$};
   \draw[<-] (h2) -- ++(-5mm,0) node[pos=0.5,anchor=south] {$h[1]$};
   \draw[<-] (h3) -- ++(-5mm,0) node[pos=0.5,anchor=south] {$h[2]$};
   \draw node[sysadd, below=of h2.south,yshift=5mm] (h) {$+$};
   \draw[fill] (u4) circle(.5pt) -- (h1) edge[->] (h)
               (u5) circle(.5pt) -- (h2) edge[->] (h)
               (u6) circle(.5pt) -- (h3) edge[->] (h);
   \draw[-o] (h) -- ++(5mm,0) node[right] {$r[k]$};
  \end{tikzpicture}\vspace{-2ex}
 \end{center}
 \caption{Concatenation of an ISI-channel with a rate-$\frac12$
          convolutional encoder $\left[g_{1,\textrm{oct}};\;g_{2,\textrm{oct}}\right] =
\left[5_\text{oct};\;7_\text{oct} \right]$ and an ISI-channel ($L=2$).}
 \label{fig:EncMapperNonCohCPM}
 \vspace*{-2ex}
\end{figure}

We first introduce convolutionally encoded PAM transmission over ISI-channels
(\cf\ example of Fig.~\ref{fig:EncMapperNonCohCPM}).
The discrete-time transmitter is composed of a rate-$\frac{K}{n}$ binary
convolutional encoder with generator polynomials $\left[\,g_i\,\right],\; 1 \leq
i \leq n$, with $K$ input symbols and $n$ parallel output symbols at each time
instant, a mapper and $M$-ary PAM transmission. The transmit signal traverses
through a memory-$L$ discrete-time ISI-channel with $L+1$ channel coefficients
$h[k]$ with $k$ denoting the time index. In the convolutional encoder 
\tikz\draw node[baseline,sysadd,rectangle,inner xsep=-.3pt,inner ysep=-.2pt] (g1) {$+$};
symbolizes the addition operation over the Galois field $\mathbb{F}_2$, \ie,
calculations are performed $\operatorname{mod}2$, whereas 
\tikz\draw node[baseline,sysadd] {$+$}; and
\tikz\draw node[baseline,sysmul] {$\times$};
indicate the addition and multiplication operation over the real numbers,
respectively.

\section{Matched Encoding Approach}\label{sec:matchedEncDerivation}

In the conventional approach one would process the receiver input signal first
by a MLSE or a symbol-by-symbol trellis-based equalizer for the FIR filter $h[k]$ and
forward soft- or hard-output symbols of this trellis equalization to the
decoder for the channel code, \ie, solve the equalization and decoding tasks in
two separate processing steps. An optimum receiver however would perform MLSE
in the super-trellis, decoding the binary channel encoder and the ISI-channel
impulse response $h[k]$ of length $L$ jointly. In a straight-forward approach
the super-trellis would have $Z_\mathrm{enc}\cdot M^L$ states, where
$Z_\mathrm{enc}$ is the number of states of the convolutional encoder.

In order to reduce the computational complexity of STD, we introduce a matched
trellis description for convolutionally encoded PAM transmission over
ISI-channels (so-called matched encoding). This non-linear trellis encoder can
be used to build the matched decoding (MD) trellis for the joint equalization
and decoding process.

If the number of output symbols from the encoder is related to the size of the
modulation alphabet $M$ so that $n=\log_2(M)$ holds, the following trellis
description can achieve exactly the same performance at reduced complexity,
\ie, with fewer states in the trellis. To see this, note that in each encoding
step, $n-K$ output symbols of the encoder are redundant and depend on $K$ input
symbols. \Eg, in Fig.~\ref{fig:EncMapperNonCohCPM} one of the two channel
encoder output symbols contains no further information.

The restriction that the size of the modulation alphabet has to match the
number of output symbols of the convolutional encoder only allows to combine a
$\frac{K}{2}$-rate encoder with a $4$-ary modulation, a $\frac{K}{3}$-rate
encoder with a $8$-ary modulation, and so on. However, we showed
in~\cite{2012arXiv1208.0193S} that when puncturing is performed to increasing
the rate of the convolutional encoder the matched decoding approach can still
be applied using a time-variant non-linear trellis description and a
slightly modified VA.

We here show, how to merge the binary channel encoder with the $M$-ary channel
impulse response to form a single time-invariant binary non-linear trellis
encoder. To this end, we transform the transmission scheme step-by-step.

First, we describe the mapping process analytically.\footnote{To simplify
notation, we do no longer strictly distinguish elements and operations from the
Galois field and the real numbers.} For clarity, we restrict ourselves to
$M=4$, but note that the concept easily extends to arbitrary $M=2^n$.

In this example $M=4$, \ie, $n=2$, with natural mapping (here equals a set
partitioning mapping), the upper branch corresponds to the most significant bit
(MSB) whereas the lower branch describes the least significant bit (LSB).
Having the MSB and LSB at time instant $k$ we now have to perform the mapping
to the symbols of the modulation alphabet.

\begin{figure}[ht]\vspace{-2ex}
 \begin{center}
  \begin{tikzpicture}[>=latex,x=1em,y=4ex,font=\footnotesize,inner sep=0.3em,
                      node distance=10mm and 4mm]
   \node at (0,0) (u) {$u[k]$};
   \node[coordinate,right=of u] (in) {};
   \draw node[syslinear,right=of in] (T1) {};
   \draw node[syslinear,right=of T1] (T2) {};
   \draw node[sysadd, above=of T1.east,xshift=2mm,rectangle,inner xsep=-.3pt,inner ysep=-.2pt] (g1) {$+$};
   \draw node[sysadd, below=of T1.east,xshift=2mm,rectangle,inner xsep=-.3pt,inner ysep=-.2pt] (g2) {$+$};
   \node[anchor=south,at=(g1.north)] {$\operatorname{mod}2$};
   \node[anchor=north,at=(g2.south)] {$\operatorname{mod}2$};
   \path (u)       edge[o->] node[coordinate,pos=0.6] (u1) {} (T1)
         (T1)      edge[->]  node[coordinate,midway] (u2) {} (T2)
         (T2.east) edge[-]   node[coordinate,pos=1] (u3) {}  ++(2.5mm,0);
   \draw[fill]  (u1) circle(.5pt) -- ++(0,5mm) edge[->] (g1)
                (u3) circle(.5pt) -- ++(0,5mm) edge[->] (g1)
                (u1) circle(.5pt) -- ++(0,-5mm) edge[->] (g2)
                (u2) circle(.5pt) -- ++(0,-5mm) edge[->] (g2)
                (u3) circle(.5pt) -- ++(0,-5mm) edge[->] (g2);
   \node[sysmul,at=(g1.east),xshift=12.5mm] (g1times2) {$\times$};
   \draw[<-] (g1times2) -- ++(0,5mm) node[anchor=south] {$2$};
   \node[sysadd,at=(T2),xshift=15mm] (g1plusg2) {$+$};
   \draw[->] (g1) -- (g1times2) -| (g1plusg2);
   \draw[->] (g2) -| (g1plusg2);
   \node[sysmul,right=of g1plusg2] (bipolar1) {$\times$};
   \draw[<-] (bipolar1) -- ++(0,5mm) node[anchor=south] {$2$};
   \node[sysadd,right=of bipolar1] (bipolar2) {$+$};
   \draw[<-] (bipolar2) -- ++(0,7.5mm) node[anchor=south] {$-(M-1)$};
   \draw node[syslinear,at=(bipolar2),xshift=10mm] (T3) {};
   \draw node[syslinear,right=of T3] (T4) {};
   \draw[-] (g1plusg2) -- (bipolar1) -- (bipolar2);
   \path (bipolar2) edge[->] node[coordinate,midway] (u4) {} (T3)
         (T3)       edge[->] node[coordinate,midway] (u5) {} (T4)
         (T4.east)  edge[-] node[coordinate,pos=1]   (u6) {} ++(2.5mm,0);
   \node[sysmul,at=(u4),yshift=-10mm] (h1) {$\times$};
   \node[sysmul,at=(u5),yshift=-10mm] (h2) {$\times$};
   \node[sysmul,at=(u6),yshift=-10mm] (h3) {$\times$};
   \draw[<-] (h1) -- ++(-5mm,0) node[pos=0.5,anchor=south] {$h[0]$};
   \draw[<-] (h2) -- ++(-5mm,0) node[pos=0.5,anchor=south] {$h[1]$};
   \draw[<-] (h3) -- ++(-5mm,0) node[pos=0.5,anchor=south] {$h[2]$};
   \draw node[sysadd, below=of h2.south,yshift=5mm] (h) {$+$};
   \draw[fill] (u4) circle(.5pt) -- (h1) edge[->] (h)
               (u5) circle(.5pt) -- (h2) edge[->] (h)
               (u6) circle(.5pt) -- (h3) edge[->] (h);
   \draw[-o] (h) -- ++(5mm,0) node[right] {$r[k]$};
  \end{tikzpicture}\vspace{-1ex}
 \end{center}
 \caption{Equivalent description of the convolutional encoding and the ISI-channel
          (exemplarily $M=4$; natural labeling).}
 \label{fig:EncStep2NonCohCPM}
 \vspace*{-1ex}
\end{figure}
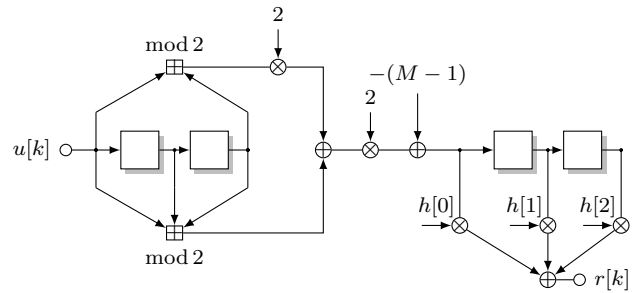

For the $4$-ary natural labeling we multiply the MSB by $2$ and add the LSB,
\ie, $c[k]=2\text{MSB}[k]+\text{LSB}[k]$. The conversion from unipolar binary
symbols $c[k]$ into bipolar symbols $b[k]$ within an alphabet of size $M$ can
be done with $b[k] = (c[k]\cdot2)-(M-1)$.The resulting block diagram, for
natural labeling is depicted in Fig.~\ref{fig:EncStep2NonCohCPM}.

Different mappings are easily incorporated, \eg, a Gray labeling can be achieved
by $c[k] = (1-\text{MSB}[k])(2\text{MSB}[k] + \text{LSB}[k]) + (\text{MSB}[k])
(2 \text{MSB}[k] + (1-\text{LSB}[k]))$. Furthermore, a $4$-ary quadrature
amplitude modulation (QAM) can be represented using the MSB as real part and
the LSB as imaginary part (or vice versa), \eg, $b[k] = (2\text{MSB}-1) +
\imag(2\text{LSB}-1)$, with $\imag=\sqrt{-1}$. 

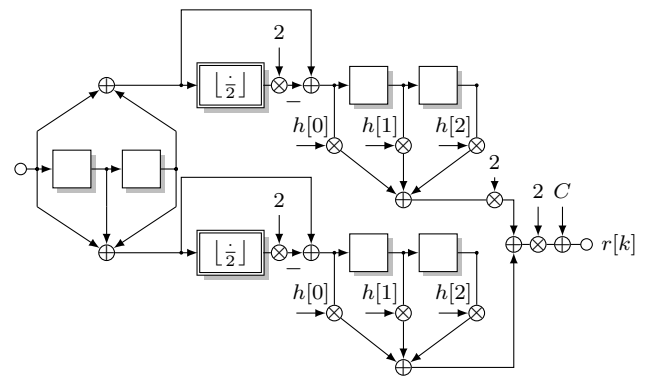
\begin{figure}[ht]\vspace{-1ex}
 \begin{center}
  \begin{tikzpicture}[>=latex,x=10em,y=4ex,font=\footnotesize,inner sep=0.3em,
                      node distance=10mm and 4mm]
    \node at (0,0) (u) {};
    \node[coordinate,right=of u] (in) {};
    \draw node[syslinear,right=of in,xshift=-3mm] (T1) {};
    \draw node[syslinear,right=of T1] (T2) {};
    \draw node[sysadd, above=of T1.east,xshift=2mm] (g1) {$+$};
    \draw node[sysadd, below=of T1.east,xshift=2mm] (g2) {$+$};
    \path (u)       edge[o->] node[coordinate,pos=0.6] (u1) {} (T1)
          (T1)      edge[->]  node[coordinate,midway] (u2) {} (T2)
          (T2.east) edge[-]   node[coordinate,pos=1] (u3) {}  ++(2.0mm,0);
    \draw[fill]  (u1) circle(.5pt) -- ++(0,5mm) edge[->] (g1)
                 (u3) circle(.5pt) -- ++(0,5mm) edge[->] (g1)
                 (u1) circle(.5pt) -- ++(0,-5mm) edge[->] (g2)
                 (u2) circle(.5pt) -- ++(0,-5mm) edge[->] (g2)
                 (u3) circle(.5pt) -- ++(0,-5mm) edge[->] (g2);
    \node[sysnonlinear,at=(g1),xshift=12mm] (floorMSB) {$\left\lfloor\frac{\cdot}{2}\right\rfloor$};
    \draw[->] (g1) -- node[coordinate,pos=0.8] (preFloorMSB) {} (floorMSB);
    \node[sysmul,right=of floorMSB,xshift=-3mm] (MSBfloor2) {$\times$};
    \node[sysadd,right=of MSBfloor2,xshift=-2mm] (MSBfloorAdd) {$+$};
    \draw[<-] (MSBfloor2) -- ++(0,5mm) node[anchor=south] {$2$};
    \draw[->] (floorMSB) -- (MSBfloor2) -- (MSBfloorAdd) node[below left] {$-$};
    \draw[->] (preFloorMSB) -- ++(0,10mm) -| (MSBfloorAdd);
    \draw[fill] (preFloorMSB) circle(.5pt);
    \draw node[syslinear,at=(MSBfloorAdd),xshift=5mm] (T3MSB) {};
    \draw node[syslinear,right=of T3MSB] (T4MSB) {};
    \path (MSBfloorAdd) edge[->] node[coordinate,midway] (u4) {} (T3MSB)
          (T3MSB)       edge[->] node[coordinate,midway] (u5) {} (T4MSB)
          (T4MSB.east)  edge[-] node[coordinate,pos=1]   (u6) {} ++(2.5mm,0);
    \node[sysmul,at=(u4),yshift=-8mm] (MSBh1) {$\times$};
    \node[sysmul,at=(u5),yshift=-8mm] (MSBh2) {$\times$};
    \node[sysmul,at=(u6),yshift=-8mm] (MSBh3) {$\times$};
    \draw[<-] (MSBh1) -- ++(-5mm,0) node[pos=0.5,anchor=south] {$h[0]$};
    \draw[<-] (MSBh2) -- ++(-5mm,0) node[pos=0.5,anchor=south] {$h[1]$};
    \draw[<-] (MSBh3) -- ++(-5mm,0) node[pos=0.5,anchor=south] {$h[2]$};
    \draw node[sysadd, below=of MSBh2.south,yshift=5mm] (MSBh) {$+$};
    \draw[fill] (u4) circle(.5pt) -- (MSBh1) edge[->] (MSBh)
                (u5) circle(.5pt) -- (MSBh2) edge[->] (MSBh)
                (u6) circle(.5pt) -- (MSBh3) edge[->] (MSBh);
    \node[sysmul,at=(MSBh),xshift=12mm] (MSBh2) {$\times$};
    \draw[<-] (MSBh) -- (MSBh2) -- ++(0,3mm) node[anchor=south] {$2$};
    \node[sysnonlinear,at=(g2),xshift=12mm] (floorLSB) {$\left\lfloor\frac{\cdot}{2}\right\rfloor$};
    \draw[->] (g2) -- node[coordinate,pos=0.8] (preFloorLSB) {} (floorLSB);
    \node[sysmul,right=of floorLSB,xshift=-3mm] (LSBfloor2) {$\times$};
    \node[sysadd,right=of LSBfloor2,xshift=-2mm] (LSBfloorAdd) {$+$};
    \draw[<-] (LSBfloor2) -- ++(0,5mm) node[anchor=south] {$2$};
    \draw[->] (floorLSB) -- (LSBfloor2) -- (LSBfloorAdd) node[below left] {$-$};
    \draw[->] (preFloorLSB) -- ++(0,10mm) -| (LSBfloorAdd);
    \draw[fill] (preFloorLSB) circle(.5pt);
    \draw node[syslinear,at=(LSBfloorAdd),xshift=5mm] (T3LSB) {};
    \draw node[syslinear,right=of T3LSB] (T4LSB) {};
    \path (LSBfloorAdd) edge[->] node[coordinate,midway] (u4) {} (T3LSB)
          (T3LSB)       edge[->] node[coordinate,midway] (u5) {} (T4LSB)
          (T4LSB.east)  edge[-] node[coordinate,pos=1]   (u6) {} ++(2.5mm,0);
    \node[sysmul,at=(u4),yshift=-8mm] (LSBh1) {$\times$};
    \node[sysmul,at=(u5),yshift=-8mm] (LSBh2) {$\times$};
    \node[sysmul,at=(u6),yshift=-8mm] (LSBh3) {$\times$};
    \draw[<-] (LSBh1) -- ++(-5mm,0) node[pos=0.5,anchor=south] {$h[0]$};
    \draw[<-] (LSBh2) -- ++(-5mm,0) node[pos=0.5,anchor=south] {$h[1]$};
    \draw[<-] (LSBh3) -- ++(-5mm,0) node[pos=0.5,anchor=south] {$h[2]$};
    \draw node[sysadd, below=of LSBh2.south,yshift=5mm] (LSBh) {$+$};
    \draw[fill] (u4) circle(.5pt) -- (LSBh1) edge[->] (LSBh)
                (u5) circle(.5pt) -- (LSBh2) edge[->] (LSBh)
                (u6) circle(.5pt) -- (LSBh3) edge[->] (LSBh);
    \node[sysadd,at=(T2),xshift=49mm,yshift=-10mm] (sumAll) {$+$};
    \draw[->] (LSBh) -| (sumAll);
    \draw[->] (MSBh2) -| (sumAll);
    \node[sysmul,right=of sumAll,xshift=-3mm] (alltimes) {$\times$};
    \node[sysadd,right=of alltimes,xshift=-3mm] (allplus)  {$+$};
    \draw[<-] (alltimes) -- ++(0,5mm) node[anchor=south] {$2$};
    \draw[<-] (allplus) -- ++(0,5mm) node[anchor=south] {$C$};
    \draw[-o] (sumAll) -- (alltimes) -- (allplus) -- ++(4mm,0) node[right] {$r[k]$};
  \end{tikzpicture}\vspace{-1ex}
 \end{center}
 \caption{Replacement of the $\operatorname{mod}2$ addition with the
          non-linear representation using the $\operatorname{floor}$ function.}
 \label{fig:EncStep3NonCohCPM}
 \vspace*{-1ex}
\end{figure}

For the second step, recall that the $\operatorname{mod}$-operation can be
represented using the $\operatorname{floor}$-function. In terms of Gaussian
notation we can thus write
\begin{equation}
 x\operatorname{mod} n = x - n\cdot \left\lfloor \frac{x}{n} \right\rfloor
\end{equation}
where $\left\lfloor . \right\rfloor$ denotes the \texttt{floor}-function.
In addition, we see that the main branch (after the summation of MSB and LSB)
has a multiplication and summation which can be moved to the output of the
convolution. With $C = -\sum\nolimits_{k=0}^{L}h[k](M-1)$ and the Gauss
representation of the modulo operation we can sketch the transmission system as
depicted in Fig.~\ref{fig:EncStep3NonCohCPM}. All calculations can therefore be
performed in the real numbers.

Finally, note that now the convolution can be moved into the MSB branch and LSB
branch, respectively, which enables to use binary delay elements instead of
$M$-ary ones. The mapping can be moved to the end of the branches.

This representation now has $n$ independent binary branches, \ie, an MSB and an
LSB branch in the case of $n=2$, which all depend on the same $K$ input values
(here $K=1$). Due to the memory elements of the ISI-channel being binary and
depending on either the MSB or the LSB we can combine them with the memory
elements of the convolutional encoder and distinguish them using the generator
polynomials $g_1$ and $g_2$. This results in a single non-linear filter
combining the calculations in each branch.

\begin{figure}[ht]\vspace{-1ex}
 \begin{center}
  \newcommand{\EQMSB}{$\displaystyle
    \left( u[k]*g_1[k] \right)*h[k]
  - 2\cdot\left\lfloor \frac{u[k]*g_1[k]}{2}\right\rfloor * h[k]$}
  \newcommand{\EQLSB}{$\displaystyle
    \left( u[k]*g_2[k] \right)*h[k]
  - 2\cdot\left\lfloor \frac{u[k]*g_2[k]}{2}\right\rfloor * h[k]$}
  \begin{tikzpicture}[>=latex,x=10em,y=4ex,font=\footnotesize,inner sep=0.3em,
                      node distance=10mm and 10mm]
   \node (u) {$u[k]$};
   \node[coordinate,right=of u,xshift=-12mm] (in) {};
   \draw node[syslinear,right=of in] (T1) {};
   \draw node[syslinear,right=of T1] (T2) {};
   \draw node[syslinear,right=of T2] (T3) {};
   \draw node[syslinear,right=of T3] (T4) {};
   \path (u)       edge[o->] node[coordinate,pos=0.6] (u1) {} (T1)
         (T1)      edge[->]  node[coordinate,midway]   (u2) {} (T2)
         (T2)      edge[->]  node[coordinate,midway]   (u3) {} (T3)
         (T3)      edge[->]  node[coordinate,midway]   (u4) {} (T4)
         (T4.east) edge[-]   node[coordinate,pos=1]    (u5) {}  ++(2.5mm,0);
   \node[sysnonlinear,minimum width=45mm,anchor=south west,at=(T1.west),xshift=-5mm,yshift=5mm]  (MSB) {\EQMSB};
   \node[sysnonlinear,minimum width=45mm,anchor=north west,at=(T1.west),xshift=-5mm,yshift=-5mm] (LSB) {\EQLSB};
   \draw[fill] (u1) circle(.5pt) edge[->] ++(0,5mm)
               (u1) circle(.5pt) edge[->] ++(0,-5mm)
               (u2) circle(.5pt) edge[->] ++(0,5mm)
               (u2) circle(.5pt) edge[->] ++(0,-5mm)
               (u3) circle(.5pt) edge[->] ++(0,5mm)
               (u3) circle(.5pt) edge[->] ++(0,-5mm)
               (u4) circle(.5pt) edge[->] ++(0,5mm)
               (u4) circle(.5pt) edge[->] ++(0,-5mm)
               (u5) circle(.5pt) edge[->] ++(0,5mm)
               (u5) circle(.5pt) edge[->] ++(0,-5mm);
   \node[sysmul,right=of MSB,xshift=-9mm] (mul2MSB) {$\times$};
   \draw[<-] (mul2MSB) -- ++(0,5mm) node[anchor=south] {$2$};
   \node[sysadd,at=(T4),xshift=13mm] (sumAll) {$+$};
   \draw[->] (LSB) -| (sumAll);
   \draw[->] (MSB) -- (mul2MSB) -| (sumAll);
   \node[sysmul,right=of sumAll,xshift=-9mm] (alltimes) {$\times$};
   \node[sysadd,right=of alltimes,xshift=-9mm] (allplus)  {$+$};
   \draw[<-] (alltimes) -- ++(0,5mm) node[anchor=south] {$2$};
   \draw[<-] (allplus)  -- ++(0,-5mm) node[anchor=north] {$C$};
   \draw[-o] (sumAll) -- (alltimes) -- (allplus) -- ++(5mm,0) node[above] {$r[k]$};
  \end{tikzpicture}\vspace{0ex}
 \end{center}
 \caption{The matched encoder (ME) as a non-linear encoder representation of
          coded PAM transmission over an ISI-channel.}
 \label{fig:EncStep4NonCohCPM}
 \vspace*{-1ex}
\end{figure}
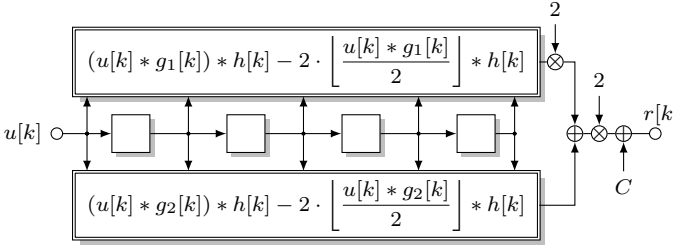
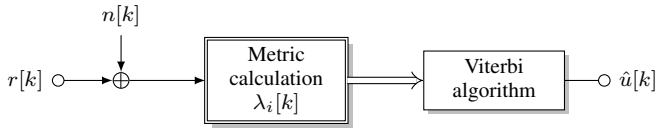
\begin{figure}[ht]\vspace{-2ex}
 \begin{center}
  \begin{tikzpicture}[>=latex,x=10em,y=4ex,font=\footnotesize,inner sep=0.3em,
                      node distance=10mm and 10mm]
   \node (u) {$r[k]$};
   \node[coordinate,right=of u,xshift=-12mm] (in) {};
   \draw node[sysadd,right=of in] (pNoise) {$+$};
   \draw node[sysnonlinear, right=of pNoise] (metric) {\parbox{1.5cm}{\centering Metric calculation\\ $\lambda_i[k]$}};
   \draw node[syslinear, right=of metric] (VA) {\parbox{1.5cm}{\centering Viterbi algorithm}};
   \draw[<-] (pNoise) -- ++(0,6mm) node[above] {$n[k]$};
   \draw[o->] (u.east) -- (pNoise);
   \draw[->] (pNoise) -- (metric);
   \draw[double,double distance=2pt,-implies] (metric) -- (VA);
   \draw[-o] (VA.east) -- ++(6mm,0) node[right] {$\hat{u}[k]$};
  \end{tikzpicture}\vspace{0ex}
 \end{center}
 \caption{The full-state matched decoder (MD) using the non-linear encoder
          representation for metric calculation and the VA for decoding.}
 \label{fig:MDMetricVA}
 \vspace*{-1ex}
\end{figure}

The resulting, non-linear trellis encoder, as depicted in
Fig.~\ref{fig:EncStep4NonCohCPM}, can be used to generate the hypothesis and
the state transitions of a finite state machine (FSM). The receiver is depicted
in Fig.~\ref{fig:MDMetricVA} and uses the hypothesis to calculate the metrics
$\lambda_i[k]$, \eg, Euclidean distances, for the noisy received signal and
performs optimum MLSE via the VA. At this point a suboptimum state reduction
can be applied, as described in Sec.~\ref{sec:RSSE}.

\section{Complexity Comparison}\label{sec:complexity}

The main advantage of matched encoding is the reduction of the convolution by
the ISI-channel from an $M$-ary input sequence into $\log_2(M)$ binary parallel
convolutions in each branch. As the number of convolutions affects the
calculation of metrics at the receiver but does not influence the number of
resulting MLSE states, we will now examine the complexity of separated
equalization and decoding, the super-trellis decoding (STD), and matched
decoding (MD). Clearly, as a measure for the computational complexity the total
number of states required for receiver-side processing can be adopted. For our
comparison we need to distinguish the number of states that result from
the convolutional encoder and the ISI-channel from the receiver complexity. The
latter can either be a result of separated equalization and decoding,
super-trellis decoding, or matched decoding.


\subsection{Separated Equalization and Decoding}

For separated equalization and decoding the receiver complexity is defined as the
sum of states in the equalization and the decoding, \ie,
\begin{align}
 Z_\mathrm{separate}= Z_\mathrm{equ} + Z_\mathrm{enc}.
\end{align}
In our simulations we distinguish between hard- and soft-output trellis-based
equalization using DFSE, or the BCJR algorithm, respectively, and decoding is
performed using the VA in the full-state trellis.

\subsection{Super-Trellis Decoding}

In a super-trellis we consider encoder states and channel states
jointly resulting in a total number of states in the super-trellis of
\begin{align}
 Z_{\mathrm{STD}} = Z_{\mathrm{enc}}\cdot Z_{\mathrm{equ}} 
               = 2^\nu \cdot M^L = 2^\nu \cdot 2^{(n\cdot L)}.
\end{align}
Apparently, already for moderate $\nu$, $n$ and/or $L$, super-trellis decoding
becomes intractable.

\subsection{Matched Decoding}

There are two differences compared to STD when considering the proposed matched
encoding/decoding approach. First, the convolution with the channel impulse
response is done with binary delay elements in contrast to $M$-ary elements.
Second, as the MSB and LSB depend on each other (as of the channel encoder) not
all state transitions are allowed anymore. As can be seen from
Fig.~\ref{fig:EncStep4NonCohCPM} the total number of delay elements does not
increase although we use binary delay elements, only. Thus, we still have
$2^\nu$ possible states for the binary channel encoder (which is fully
integrated into the non-linear encoder) but only $2^L$ possible states for the
convolution resulting in a total number of states of
\begin{align}
 Z_{\mathrm{MD}} &= 2^\nu \cdot 2^{L}.
\end{align}
Recall that for $n=2$ there are two convolutions in parallel for the
computation of the hypothesis. Finally, employing RSSE (\cf\
Sec.~\ref{sec:RSSE}), the complexity depends on the partitioning as will be
described below, \ie, $Z_\text{MD-RSSE} = Z_\text{R} = 2^r$ with arbitrary
integer $r>1$.
\subsection{Comparison}
\begin{table}
 \begin{center}
  \caption{Number of states for PAM transmission with $M=4$, $n=2$
           and for the super-trellis representation and MD, respectively.}
  \begin{tabular}[ht]{ccccc}\toprule
   Encoder                                                           & $L$ & $Z_{\mathrm{STD}}$ & $Z_{\mathrm{MD}}$ & $G_{\text{MD}}$ \\\midrule
   \multirowbt{6}{*}{
   	  \parbox{2.5cm}{\begin{center}$16$ states\\$\nu=4$\\\eg, $\left[23_\text{oct};\;04_\text{oct}\right]$\end{center}}
   }
                                                                     & $0$ & $16$    & $16$   & $1$             \\\cmidrule{2-5}
                                                                     & $1$ & $64$    & $32$   & $2$             \\\cmidrule{2-5}
                                                                     & $2$ & $256$   & $64$   & $4$             \\\cmidrule{2-5}
                                                                     & $3$ & $1024$  & $128$  & $8$             \\\cmidrule{2-5}
                                                                     & $4$ & $4096$  & $256$  & $16$            \\\cmidrule{2-5}
                                                                     & $5$ & $16384$ & $512$  & $32$            \\
   \midrule
   \multirowbt{6}{*}{
   	  \parbox{2.5cm}{\begin{center}$64$ states\\$\nu=6$\\\eg, $\left[103_\text{oct};\;024_\text{oct}\right]$\end{center}}
   }
                                                                     & $1$ & $64$    & $64$   & $1$             \\\cmidrule{2-5}
                                                                     & $0$ & $256$   & $128$  & $2$             \\\cmidrule{2-5}
                                                                     & $2$ & $1024$  & $256$  & $4$             \\\cmidrule{2-5}
                                                                     & $3$ & $4096$  & $512$  & $8$             \\\cmidrule{2-5}
                                                                     & $4$ & $16384$ & $1024$ & $16$            \\\cmidrule{2-5}
                                                                     & $5$ & $65536$ & $2048$ & $32$            \\
      \bottomrule
   \label{tab:gainStates}\vspace*{-5ex}
  \end{tabular}
 \end{center}
\end{table}

The main advantage of MD compared to STD is the reduction of states without
loss in performance. The resulting trellis still describes the super-trellis
but with fewer states. The gain of this state reduction therefore calculates to
\begin{align}
 G_{\text{MD}} = \frac{Z_{\text{STD}}}{Z_{\text{MD}}}
               = \frac{2^{(n\cdot L)}}{2^{L}}
               = 2^{L(n - 1)}.
\end{align}
Table~\ref{tab:gainStates} summarizes several examples for different encoders
and channel lengths for the special case of $n=2$ \mbox{($M=4$)}.
Obviously the gain increases with the length of the ISI-channel.

\section{Reduced-State Sequence Estimation}\label{sec:RSSE}

We have shown that the super-trellis of convolutionally encoded transmission
over ISI-channels can be represented using significantly fewer states by
parallelizing the $M$-ary convolution. At this point we can use reduced-state
sequence estimation (RSSE)~\cite{Eyuboglu1988} to further reduce the number of
states at the cost of small loss in Euclidean distance.

\begin{figure}
 \begin{center}
  \begin{tikzpicture}[x=5cm,y=1.2cm]
   \foreach \k in {0} {
    \foreach \S/\Mpo in {0/0, 0/2,
                         1/0, 1/2,
                         2/1, 2/3,
                         3/1, 3/3
                        } {
      \draw[fill] (\k,\S) circle(1pt) -- ($(\k,\Mpo)+(1,0)$) circle (1pt);
    }
   }
  \node at (0,3) (S0) {};
  \node at (0,2) (S1) {};
  \node at (0,1) (S2) {};
  \node at (0,0) (S3) {};
  \node at (1,3) (E0) {};
  \node at (1,2) (E1) {};
  \node at (1,1) (E2) {};
  \node at (1,0) (E3) {};
  \draw (S0) ++(0,0.4cm) arc(90: -90:0.4cm)
        (S0) ++(0,0.4cm) arc(90: 270:1.0cm)
                         arc(-90: 90:0.4cm)
        (S0) ++(0,-0.4cm) arc(90:270:0.2cm);
  \draw[fill=white] (S2) ++(0,0.4cm) arc(90: -90:0.4cm)
        (S2) ++(0,0.4cm) arc(90: 270:1.0cm)
                         arc(-90: 90:0.4cm)
        (S2) ++(0,-0.4cm) arc(90:270:0.2cm);
  \draw (E0) ++(0,0.4cm)  arc(90: 270:0.4cm) 
        (E0) ++(0,0.4cm)  arc(90: -90:1.0cm)
                          arc(270: 90:0.4cm)
        (E0) ++(0,-0.4cm) arc(90:-90:0.2cm);
  \draw[fill=white] (E2) ++(0,0.4cm)  arc(90: 270:0.4cm) 
        (E2) ++(0,0.4cm)  arc(90: -90:1.0cm)
                          arc(270: 90:0.4cm)
        (E3) ++(0,+0.4cm) arc(-90:90:0.2cm);
   \foreach \k in {0} {
    \foreach \S/\Mpo in {0/0, 0/2,
                         1/0, 1/2,
                         2/1, 2/3,
                         3/1, 3/3
                        } {
      \draw[fill] (\k,\S) circle(1pt) -- ($(\k,\Mpo)+(1,0)$) circle (1pt);
    }
   }
  \node[left] at (S0) {$S_0$};
  \node[left] at (S1) {$S_1$};
  \node[left] at (S2) {$S_2$};
  \node[left] at (S3) {$S_3$};
  \node[rotate=90,anchor=north] at (1.2,0.5) {Hyperstate $1$};
  \node[rotate=90,anchor=north] at (1.2,2.5) {Hyperstate $0$};
  \end{tikzpicture}\vspace*{0ex}
 \end{center}
 \caption{Illustration of hyper states and sub states in a trellis.}
 \label{fig:rsseHyperstates}
 \vspace*{0ex}
\end{figure}
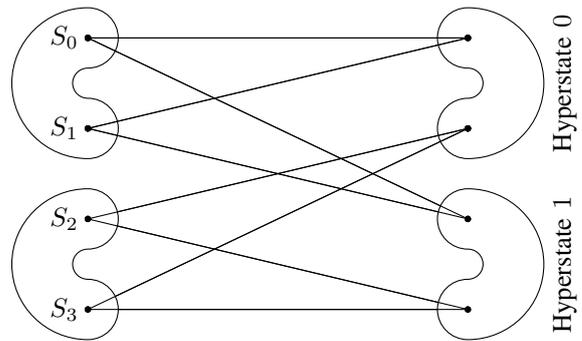

In RSSE, $Z$ MLSE states, each with $M=2^K$ possible transitions to adjacent
states, are combined into $Z_\text{R} = \frac{Z}{2^J};\;J\in\mathbb{N}$
hyperstates~\cite{huber1992trelliscodierung,Spinnler1995} each having $2^J$
substates and $2^K\cdot2^J$ state transitions as depicted in
Fig.~\ref{fig:rsseHyperstates} with $K=1$ and $J=1$. A certain assignment of
states to hyper states is called \textit{partitioning}~\cite{Spinnler1995}.

Instead of selecting a survivor from $2^K$ arriving transitions at each of the
$Z$ MLSE states we now select a single survivor from a set of $2^K\cdot2^J$
transitions at $Z_\text{R}$ hyper states. The number of metrics that have to be
calculated remains $2^K\cdot Z$.

The main difference is, that we decide for a surviving path prematurely
resulting in a truncation of error events. A loss in Euclidean distance
appears if an error event with minimum Euclidean distance gets truncated.
Therefore the performance of RSSE strongly depends on the partitioning of the
states into hyperstates.

One approach to find the optimum partitioning is to determine the mutual state
distances and iteratively maximize the intra hyperstate
distance~\cite{Spinnler1995}. Unfortunately the exhaustive search for the state
distances is impractical for a larger number of states.

Fortunately the channel impulse response $h[k]$ is fully integrated into the
non-linear trellis description. W.l.o.g we assume that the channel impulse
response $h[k]$ is minimum phase, which can be achieved by the application of a
proper all-pass filter.
For a minimum phase channel impulse response the MLSE equalization with a
reduced number of states is well-known as delayed decision-feedback sequence
estimation (DFSE)~\cite{Lee1977,Duel-Hallen1989,Eyuboglu1988,Eyuboglu1989}. On
an ISI-channel with delay length $L$, DFSE generates the trellis on the first
$q_h<L$ coefficients only. Some post cursors of the discrete-time impulse
response are cancelled using a decision feedback in each state using the state
register of the VA. This can be interpreted as a particular solution of RSSE
using a methodical partitioning.


As the minimum phase ISI-channel is fully integrated into the non-linear
trellis we can apply the methodical DFSE partitioning to use RSSE for MD for
convolutionally encoded PAM transmission over ISI-channels.

\section{Numerical Results}\label{sec:simresults}

The effectiveness of the approach of MD is now verified by means of numerical
simulations. We restrict ourselves to rate-$\frac12$ encoding schemes and a
$4$-ary modulation alphabet. As convolutional encoder we apply the generator
polynomials given in Table~\ref{tab:gainStates}, which, in combination with
natural labeling, result in a trellis coded modulation scheme (TCM) for $M$-ary
ASK.\footnote{Note that \cite{Alvarado2012} reveals an equivalence between TCM
encoders. There, it is shown that a $[5;\;7]$ encoder with gray labeling is
identical to a $[5;\;2]$ encoder with natural labeling.} The ISI-channel is
described by (\ref{eq:channelISI}) using $L\in\left\{ 2;\;5 \right\}$ with
$Z_\text{cha} = 16$ states and $1024$ states, respectively.
For simplicity an exemplary minimum phase ISI-channel is generated by 
\begin{align}\label{eq:channelISI}
 h[k]     & = \frac{1}{\alpha}\;\cdot\;\frac{L-k+1}{L+1}; \qquad 0\leq k\leq L\\
 \alpha^2 & = \sum\limits_{k=0}^L\left( \frac{L-k+1}{L+1} \right)^2
\end{align}
and normalized to unit energy. Please note that due to the normalization the
equivalent energy per bit $E_\text{b}$ is identical at transmitter output and
receiver input.

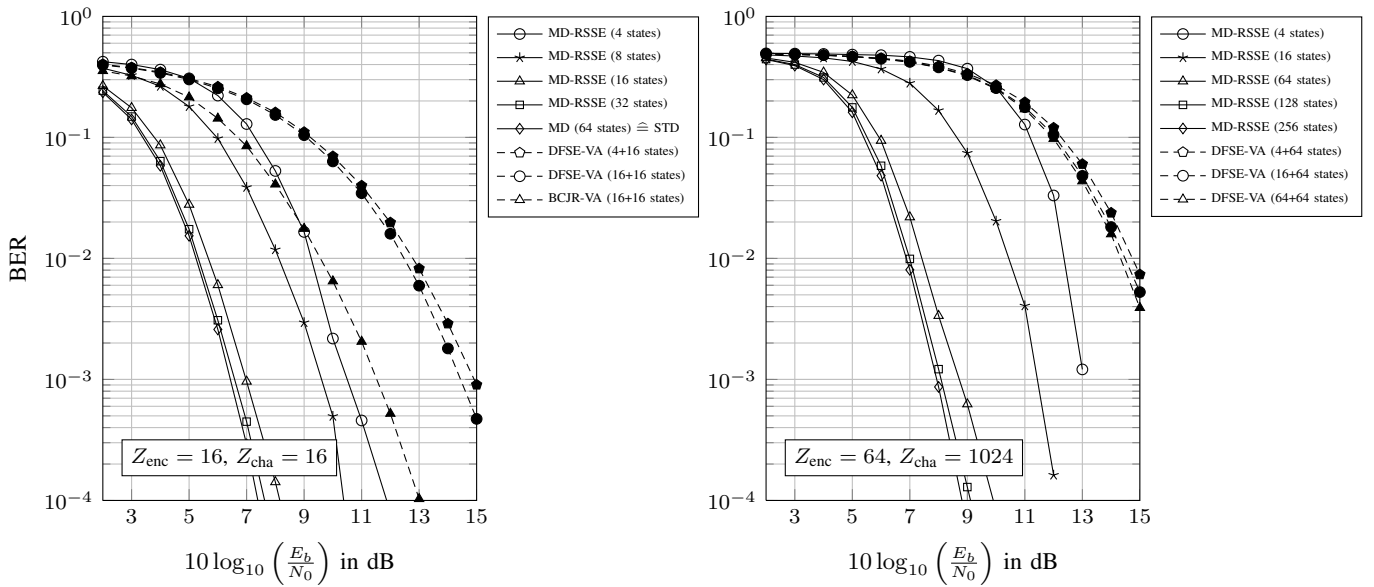
\begin{figure*}[ht]
 \begin{subfigure}[t]{0.48\textwidth}
  \begin{center}
   \begin{tikzpicture}
    \begin{axis}[
                 width=6.5cm,
                 height=8cm,
                 xlabel={$10\log_{10}\left(\frac{E_b}{N_0}\right)$ in dB},
                 ylabel={BER},
                 enlargelimits=false,
                 ymode=log,
                 cycle list name=colorsBERMDDFSE,
                 grid=both,
                 xmin=2,xmax=15,
                 ymax=1,ymin=1e-4,
                 legend pos=outer north east,
                 every axis legend/.append style={font={\tiny},nodes={right}},
                 xtick={1,3,...,15},
                ]
     \addplot table[x index=0,y index=3]          {data_pam_MD_natural256-Senc16-Sfir16-Coded.data};
     \addplot table[x index=0,y index=4]          {data_pam_MD_natural256-Senc16-Sfir16-Coded.data};
     \addplot table[x index=0,y index=5]          {data_pam_MD_natural256-Senc16-Sfir16-Coded.data};
     \addplot table[x index=0,y index=6]          {data_pam_MD_natural256-Senc16-Sfir16-Coded.data};
     \addplot table[x index=0,y index=7]          {data_pam_MD_natural256-Senc16-Sfir16-Coded.data};
     \addplot table[x index=0,y index=8]          {data_pam_MD_natural256-Senc16-Sfir16-Coded.data};
     \addplot table[x index=0,y index=9]          {data_pam_MD_natural256-Senc16-Sfir16-Coded.data};
     \addplot table[x index=0,y index=10]         {data_pam_MD_natural256-Senc16-Sfir16-Coded.data};
     \addlegendentry{MD-RSSE (4 states)}
     \addlegendentry{MD-RSSE (8 states)}
     \addlegendentry{MD-RSSE (16 states)}
     \addlegendentry{MD-RSSE (32 states)}
     \addlegendentry{MD (64 states) $\entspricht$ STD}
     \addlegendentry{DFSE-VA (4+16 states)}
     \addlegendentry{DFSE-VA (16+16 states)}
     \addlegendentry{BCJR-VA (16+16 states)}
     \coordinate (note) at (rel axis cs:0.05,0.05);
    \end{axis}
    \node[draw,fill=white,anchor=south west,font={\footnotesize}] at (note) {$Z_\text{enc} = 16$, $Z_\text{cha}=16$};
   \end{tikzpicture}
  \end{center}
 \end{subfigure}%
 \hspace{1em}
 \begin{subfigure}[t]{0.48\textwidth}
  \begin{center}
   \begin{tikzpicture}
    \begin{axis}[
                 width=6.5cm,
                 height=8cm,
                 xlabel={$10\log_{10}\left(\frac{E_b}{N_0}\right)$ in dB},
                 enlargelimits=false,
                 ymode=log,
                 cycle list name=colorsBERMDDFSE,
                 grid=both,
                 xmin=2,xmax=15,
                 ymax=1,ymin=1e-4,
                 legend pos=outer north east,
                 every axis legend/.append style={font={\tiny},nodes={right}},
                 xtick={1,3,...,15},
                ]
     \addplot table[x index=0,y index=3]          {data_pam_MD_natural65536-Senc64-Sfir1024-Coded.data};
     \addplot table[x index=0,y index=5]          {data_pam_MD_natural65536-Senc64-Sfir1024-Coded.data};
     \addplot table[x index=0,y index=7]          {data_pam_MD_natural65536-Senc64-Sfir1024-Coded.data};
     \addplot table[x index=0,y index=8]          {data_pam_MD_natural65536-Senc64-Sfir1024-Coded.data};
     \addplot table[x index=0,y index=9]          {data_pam_MD_natural65536-Senc64-Sfir1024-Coded.data};
     \addplot table[x index=0,y index=10]         {data_pam_MD_natural65536-Senc64-Sfir1024-Coded.data};
     \addplot table[x index=0,y index=11]         {data_pam_MD_natural65536-Senc64-Sfir1024-Coded.data};
     \addplot table[x index=0,y index=12]         {data_pam_MD_natural65536-Senc64-Sfir1024-Coded.data};
     \addlegendentry{MD-RSSE (4 states)}
     \addlegendentry{MD-RSSE (16 states)}
     \addlegendentry{MD-RSSE (64 states)}
     \addlegendentry{MD-RSSE (128 states)}
     \addlegendentry{MD-RSSE (256 states)}
     \addlegendentry{DFSE-VA (4+64 states)}
     \addlegendentry{DFSE-VA (16+64 states)}
     \addlegendentry{DFSE-VA (64+64 states)}
     \coordinate (note) at (rel axis cs:0.05,0.05);
    \end{axis}
    \node[draw,fill=white,anchor=south west,font={\footnotesize}] at (note) {$Z_\text{enc} = 64$, $Z_\text{cha}=1024$};
   \end{tikzpicture}
  \end{center}
 \end{subfigure}\vspace*{-1ex}
 \caption{Bit error performance for convolutionally encoded $4$-ASK
          transmission (natural labeling) over the ISI-AWGN-channel.
           Left: Generator polynomials $\left[ 23_\text{oct};\; 04_\text{oct}\right]$ ($Z_\text{enc}=16$), $L=2$ ($Z_\text{cha}=16$).
          Right: Generator polynomials $\left[103_\text{oct};\;024_\text{oct}\right]$ ($Z_\text{enc}=64$), $L=5$ ($Z_\text{cha}=1024$).
          \label{fig:BER}
 }
 \vspace*{-1ex}
\end{figure*}

Our MD approach of RSSE operating on the equivalent non-linear trellis
description is compared to separate equalization and decoding employing
DFSE/BCJR~\cite{BCJR74} for equalization and the full-state VA for decoding.
Here, the full-state BCJR equalization is used to compare our approach with
soft-decision equalization and decoding, whereas DFSE employs a hard-decision
reduced-state
equalization~\cite{Lee1977,Duel-Hallen1989,Eyuboglu1988,Eyuboglu1989} based on
a VA. Both equalization techniques require a full-state VA for decoding the
convolutional code.

In contrast, the MD-RSSE performes the VA on the reduced set of hyperstates and
selects the substates using delayed decisions fed back from the path register.
The applied partitioning is created equal to the methodical DFSE partitioning
and determines the number of states in the receiver trellis. 

\subsection{Bit Error Performance}

The bit error rates for transmission over the ISI-AWGN-channel (one sided power
spectral noise density $N_0$) are given in Fig.~\ref{fig:BER}. The number of
states for the DFSE/BCJR and for the VA are given in the legend, separately.
Here the number of states implemented in the receiver trellis and therefore the
receiver complexity is given directly for the MD/RSSE receiver.

Please note that by dispensing the interleaver between channel encoding and
modulation for the separated approaches, block errors that are caused by the
equalization process reduce the ability to decode due to correlated errors.
Obviously, the soft-decision separated approach results in improved bit error
rates when compared to the hard-decision separated approach. However, the
separated equalization and decoding approach is significantly outperformed by
our MD. In the case of a $16$-state convolutional encoder and a $16$-state
ISI-channel, the MD performance equals STD at $64$ MD states.

The behavior of MD-RSSE is equivalent to that of DFSE. With increasing number
of states the performance improves (especially at low number of states) and
converges to the performance of STD.

For the $64$-state encoder the performance converges slower due to
increased encoder complexity (STD not shown due to complexity constraints).

\subsection{Performance Vs. Complexity Trade-Off}

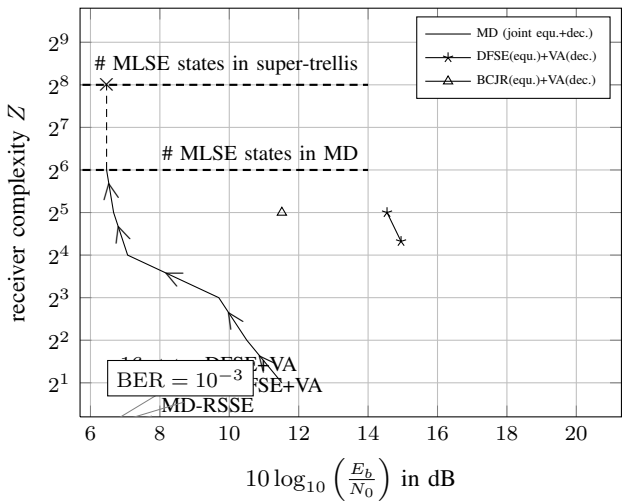
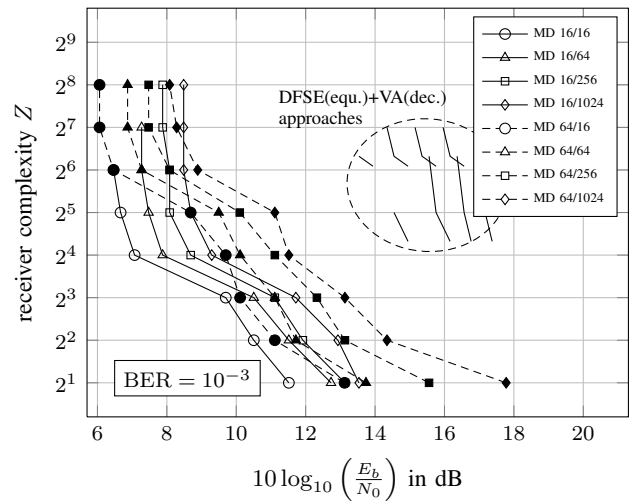
\begin{figure*}
 \begin{subfigure}[t]{0.48\textwidth}
  \begin{center}
   \begin{tikzpicture}
    \begin{axis}[
                 width=\textwidth,
                 height=7cm,
                 ylabel={receiver complexity $Z$},
                 xlabel={$10\log_{10}\left(\frac{E_b}{N_0}\right)$ in dB},
                 enlargelimits=true,
                 ymode=log,
                 log basis y=2,
                 ytick={2,4,8,16,32,64,128,256,512},
                 cycle list name=colors,
                 grid=both,
                 xmin=7,xmax=20,
                 ymin=2,ymax=512,
                 every axis legend/.append style={
                             font={\tiny},
                             nodes={right}},
                ]
     \addplot[black] table {data_pam_MD_natural256-Senc16-Sfir16-Coded-BER-RSSE.data}
       node[pos=0.1,pin=5:{\footnotesize{MD-RSSE}}] {};
     \path
       (axis cs:11.515152,2 ) -- node[pos=0.5,sloped,baseline] {$<$}
       (axis cs:10.505051,4 ) -- node[pos=0.5,sloped,baseline] {$<$}
       (axis cs:9.696970 ,8 ) -- node[pos=0.5,sloped,baseline] {$<$}
       (axis cs:7.070707 ,16) -- node[pos=0.5,sloped,baseline] {$<$}
       (axis cs:6.666667 ,32) -- node[pos=0.5,sloped,baseline] {$<$}
       (axis cs:6.464646 ,64);

     \draw[densely dashed,fill] (axis cs:6.464646,64) -- (axis cs:6.464646,256) node {$\times$};
     \addlegendentry{MD (joint equ.+dec.)};

     \addplot table {data_pam_MD_natural256-Senc16-Sfir16-Coded-BER-DFSE.data}
       node[pos=0,pin=30:{\footnotesize{$4$ states DFSE+VA}}] {}
       node[pos=1,pin=85:{\footnotesize{$16$ states DFSE+VA}}] {};
     \addlegendentry{DFSE(equ.)+VA(dec.)};
     \addplot coordinates { (1.1515152e+01,16+16) }
       node[pos=0,pin=280:{\footnotesize{$16$ states BCJR+VA}}] {};
     \addlegendentry{BCJR(equ.)+VA(dec.)};

     \draw[thick,densely dashed,black] (axis cs:5,256) --
                                             node[anchor=south east,pos=1]
                                             {\footnotesize{\# MLSE states in super-trellis}}
                                             (axis cs:14,256);
     \draw[thick,densely dashed,black] (axis cs:5,64) --
                                             node[anchor=south east,pos=1]
                                             {\footnotesize{\# MLSE states in MD}}
                                             (axis cs:14,64);
     \node[draw,fill=white,anchor=south west,font={\footnotesize}] at (rel axis cs:0.05,0.05) {$\mathrm{BER} = 10^{-3}$};
    \end{axis}
   \end{tikzpicture}
  \end{center}\vspace{-2ex}
  \caption{Receiver complexity for a $Z_\text{enc}=16$/$Z_\text{cha}=16$ transmission system.}
  \label{fig:NumbStatesSingleShot}
 \end{subfigure}
 \hspace{1em}
 \begin{subfigure}[t]{0.48\textwidth}
  \begin{center}
   \begin{tikzpicture}
    \begin{axis}[
                 width=\textwidth,
                 height=7cm,
                 ylabel={receiver complexity $Z$},
                 xlabel={$10\log_{10}\left(\frac{E_b}{N_0}\right)$ in dB},
                 enlargelimits=true,
                 ymode=log,
                 log basis y=2,
                 ytick={2,4,8,16,32,64,128,256,512},
                 cycle list name=colors4Empty4Full,
                 grid=both,
                 xmin=7,xmax=20,
                 ymin=2,ymax=512,
                 every axis legend/.append style={
                             font={\tiny},
                             nodes={right}},
                ]
     \addplot table {data_pam_MD_natural256-Senc16-Sfir16-Coded-BER-RSSE.data};
     \addplot table {data_pam_MD_natural1024-Senc16-Sfir64-Coded-BER-RSSE.data};
     \addplot table {data_pam_MD_natural4096-Senc16-Sfir256-Coded-BER-RSSE.data};
     \addplot table {data_pam_MD_natural16384-Senc16-Sfir1024-Coded-BER-RSSE.data};

     \addplot table {data_pam_MD_natural1024-Senc64-Sfir16-Coded-BER-RSSE.data};
     \addplot table {data_pam_MD_natural4096-Senc64-Sfir64-Coded-BER-RSSE.data};
     \addplot table {data_pam_MD_natural16384-Senc64-Sfir256-Coded-BER-RSSE.data};
     \addplot table {data_pam_MD_natural65536-Senc64-Sfir1024-Coded-BER-RSSE.data};

     \addlegendentry{MD 16/16};
     \addlegendentry{MD 16/64};
     \addlegendentry{MD 16/256};
     \addlegendentry{MD 16/1024};
     \addlegendentry{MD 64/16};
     \addlegendentry{MD 64/64};
     \addlegendentry{MD 64/256};
     \addlegendentry{MD 64/1024};

     \addplot[black] table {data_pam_MD_natural256-Senc16-Sfir16-Coded-BER-DFSE.data};
     \addplot[black] table {data_pam_MD_natural1024-Senc16-Sfir64-Coded-BER-DFSE.data};
     \addplot[black] table {data_pam_MD_natural1024-Senc64-Sfir16-Coded-BER-DFSE.data};
     \addplot[black] table {data_pam_MD_natural4096-Senc16-Sfir256-Coded-BER-DFSE.data};
     \addplot[black] table {data_pam_MD_natural4096-Senc64-Sfir64-Coded-BER-DFSE.data};
     \addplot[black] table {data_pam_MD_natural16384-Senc16-Sfir1024-Coded-BER-DFSE.data};
     \addplot[black] table {data_pam_MD_natural16384-Senc64-Sfir256-Coded-BER-DFSE.data};
     \addplot[black] table {data_pam_MD_natural65536-Senc64-Sfir1024-Coded-BER-DFSE.data};

     \node[coordinate] at (axis cs:15.5,50) (markA) {};
     \draw[densely dashed,rotate around={-5:(markA)}] (markA) ellipse (3em and 2.5em);
     \node[above=1.8em,xshift=-2em,font=\scriptsize] at (markA)
     {\parbox{2.5cm}{DFSE(equ.)+VA(dec.) approaches}};
     \node[draw,fill=white,anchor=south west,font={\footnotesize}] at (rel axis cs:0.05,0.05) {$\mathrm{BER} = 10^{-3}$};
    \end{axis}
   \end{tikzpicture}
  \end{center}\vspace{-2ex}
  \caption{Receiver complexity for two different convolutional codes ($16$ and
           $64$ states, first number in legend) and different ISI-channels
           ($Z_\text{cha}$ states, second number in legend).}
  \label{fig:NumbStatesSepara}
 \end{subfigure}
 \caption{Receiver complexity versus required SNR to guarantee a
          $\text{BER}=10^{-3}$ for separated equalization with DFSE, BCJR and
          decoding with VA compared to MD with RSSE.}
 \vspace*{-1ex}
\end{figure*}

We now compare the receiver complexity for separated equalization and decoding
to MD. We compare different channel encodings and ISI-channels defined by their
number of states in the transmitter. Channel encoders with $16$ or $64$ states
and an ISI-channel with $L=\left\{2,3,4,5\right\}$ (or
$\left\{16;\;64;\;256;\;1024\right\}$ states, respectively) are considered. The
channel encoding and the ISI-channel are described by their number of states at
the transmitter trellis, \ie, the number of states in the channel encoder and
equalizer, respectively, and abbreviated with $Z_\text{enc}/Z_\text{equ}$.

The target bit error rate is $10^{-3}$ and the receiver complexity is described
by the number of states as described in Sec.~\ref{sec:complexity}. As our
approach enables the use of RSSE, we can compare the performance for arbitrary
receiver complexity for the given target error rate. 

In Fig.~\ref{fig:NumbStatesSingleShot} the results for a convolutional encoder
with $16$ states and an ISI-channel with another $16$ states are depicted. The
super-trellis would have $256$ states. The best performance for separated
equalization and decoding is achieved with the soft-decision approach using the
(full-state) BCJR for equalization and the VA for decoding. However, the figure
also clearly shows that MD supersedes separated equalization and decoding
already for only two states in MD-RSSE.

In Fig.~\ref{fig:NumbStatesSepara} the results for multiple transmitter schemes
are depicted. Note, that all separated approaches perform worse with more
states compared to the proposed MD approach. Obviously, the $64$-state
convolutional encoder with $2^7=128$ receiver states can achieve best
performance due to increased constraint length of the convolutional encoder.
In contrast, increasing the memory of the ISI-channel reduces the minimum Euclidean
distance resulting in a degradation of the bit error performance.

It becomes also clear that a $Z_\text{enc}=16/Z_\text{cha}=16$ scheme achieves
a bit error rate of $10^{-3}$ with fewer number of receiver states, whereas a
convolutional encoder with more states, \ie, $Z_\text{enc}=64/Z_\text{cha}=16$,
achieves $10^{-3}$ with less signal-to-noise power ratio. Hence, the proposed
decoding schemes enable a flexible trade-off between complexity and
noise-robustness, \ie, power efficiency.

In summary, we conclude that it is favorable to choose a convolutional code
with a low number of states in combination with MD-RSSE, when low delay and low
complexity are required.
%
\section{Conclusion}\label{sec:conclusion}

In this paper we have shown that it is possible to reduce the number of states
for super-trellis decoding without loss in performance by transforming the
$M$-ary channel convolution into $\log_2(M)$ parallel binary convolutions.
Here, a coded ASK transmission is used, but as several other non-interleaved
transmission schemes (\eg, QAM over an ISI-channel) can be represented as a
separate channel encoder and a channel impulse response, this approach is
attractive for such schemes, as well. We showed that with MD the same
performance as super-trellis decoding can be achieved with significantly
reduced computational complexity. By using RSSE with DFSE-like partitioning
we obtain an efficient method for a trade-off between complexity and
performance. The only restriction of this approach is the strong relation
between the alphabet size of the modulation and the code rate.
 

\newpage
\IEEEtriggercmd{\enlargethispage{1in}}
\bibliographystyle{IEEEtran}
\bibliography{IEEEabrv,main,mine}
\end{document}